\documentclass[twocolumn,prb,floatfix]{revtex4}
\usepackage[]{fontenc}
\usepackage[latin1]{inputenc}
\usepackage{amsmath}
\usepackage{graphicx}
\usepackage{amssymb}

\begin{document}

\title{Self-breaking in planar few-atom Au constrictions for nm-spaced electrodes}

\author{K. O'Neill}

\email{k.oneill@med.tn.tudelft.nl}

\address{Kavli Institute of Nanoscience Delft, Delft University of Technology,
Lorentzweg 1, 2628 CJ Delft, The Netherlands}

\author{E. A. Osorio}

\address{Kavli Institute of Nanoscience Delft, Delft University of Technology,
Lorentzweg 1, 2628 CJ Delft, The Netherlands}

\author{H. S. J. van der Zant}

\address{Kavli Institute of Nanoscience Delft, Delft University of Technology,
Lorentzweg 1, 2628 CJ Delft, The Netherlands}

\begin{abstract}
We present results on electromigrated Au nanojunctions broken near
the conductance quantum $77.5\,\mu$S. At room temperature we find
that wires, initially narrowed by an actively-controlled electromigration
technique down to a few conductance quanta, continue to narrow after
removing the applied voltage. Separate electrodes form as mobile gold
atoms continuously reconfigure the constriction. We find, from results
obtained on over $300$ samples, no evidence for gold cluster formation
in junctions broken without an applied voltage, implying that gold
clusters may be avoided by using this self-breaking technique.
\end{abstract}
\maketitle
Electronic devices based on single nanometer-sized molecules show
promising routes to exploiting the functionality available through
organic synthesis. In addition such devices provide an experimental
platform to understand the electronic, ionic and mechanical degrees
of freedom of a single molecule and its coupling to the environment.
While many methods exist to create nanometer-spaced electrodes, the
presence of a gate electrode is crucial to correctly identify the
signatures of single-molecule conduction at low temperature, implying
that a planar geometry is required. In this geometry, a common method
for creating the nanometer sized electrode spacing is by electromigration,
in which a large current density of $\sim10^{8}\,$Acm$^{-2}$ is
used to deform a small gold wire until physically separated electrodes
are formed \cite{park}. It has remained a persistent challenge to
unambiguously determine the presence of a trapped single molecule
in transport measurements using the electromigration technique. Transport
measurements are hindered by the uncontrolled nature of the breaking
process, which produces nanogaps of a wide range of sizes, and the
formation of gold clusters that give signatures of Coulomb Blockade
and Kondo physics, unrelated to conduction processes through single
molecules \cite{gonzalez_sordan,heersche}. 

This paper demonstrates a `self-breaking' effect in gold wires that
are narrowed by electromigration to a few atoms. Gold nanoconstrictions
fabricated at room temperature tend to be unstable; on a time-scale
of tens of minutes or hours they break further until the conductance
reaches values $\ll100\,\mu$S without an applied voltage. Subsequent
measurements at low temperature show that self-broken wires are less
likely to produce gold clusters than samples that are actively broken
into separate electrodes. This `self-breaking' effect has also been
observed in transmission electron microscope studies \cite{emigtem},
indicating that the decreasing conductance is due to a physical separation
of the electrodes in time.

\begin{figure}
\begin{centering}\includegraphics[clip,scale=0.6]{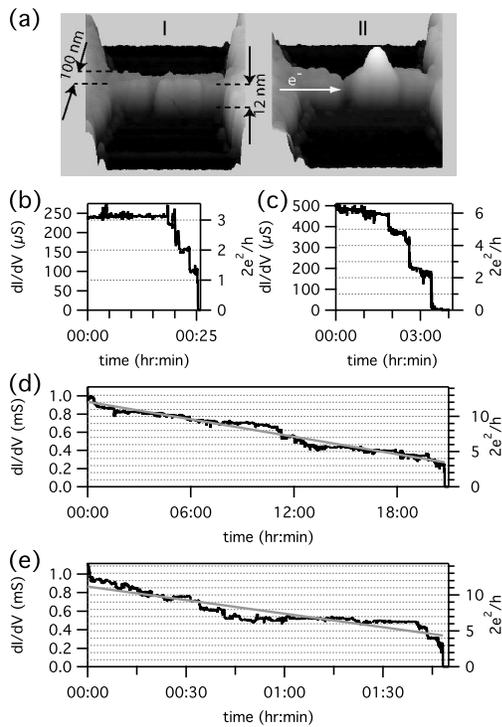}\par\end{centering}

\caption{(a) Atomic Force Microscope images of a sample before and after narrowing.
(b)-(e) show self-breaking data in which conductance (in $S$ on the
left axis and conductance quanta $\frac{2e^{2}}{h}$ on the right
axis) is plotted against time (in hours and minutes), measured at
room temperature and in a pressure less than $10^{-5}\,$Torr. The
straight lines in (d) and (e) are fits up to the points near the conductance
$2\,$G$_{0}$.}

\label{graph:fig1}
\end{figure}

Samples are fabricated as follows: thin wire bridges, with a cross-section
of $100\,$nm$ \times 12\,$nm and a typical length of $500\,$nm
are evaporated over a gate dielectric. The gates are formed using
aluminium wires and its native aluminium oxide of a few nm is used
as the gate dielectric. Contact to the bridge is made by large wires
which contribute a small series resistance \cite{trouwborst}. The
resistance between bonding pads before the wires are narrowed is typically
$100-200\,\Omega$. To initially narrow the gold wires by electromigration,
we use an active breaking scheme \cite{vanderzant}, similar to the
ones previously reported \cite{strachen_houck}. In the active breaking
process the voltage is increased from below the electromigration threshold
($>200\,$mV) while sampling the current; the maximum current is $\sim8\,$mA.
If the absolute resistance of the wire increases by a value determined
during the sweep, typically around $10$\%, the applied voltage is
reduced back to $100\,$mV, and the sweep is repeated with a new value
for the wire resistance. With this method, nanoconstrictions may be
narrowed to a target conductance with high reproducibility, often
within $10$\%, provided the target conductance is greater than the
conductance quantum G$_{0}=\frac{2e^{2}}{h}=77\,\mu$S. Electromigration
events change the junction resistance on a time-scale $<100\,\mu$s,
and the rate typically chosen for voltage output and current sampling
is $\sim25-50\,\mu$s. Figure \ref{graph:fig1}(a) shows Atomic Force
Microscope images of a sample before ($\textsf{I}$) and after ($\textsf{II}$)
narrowing using this technique. We note the formation of a hillock
downstream of the gap, demonstrating that electron wind force is responsible
for the transfer of momentum between electrons and gold atoms.

While active breaking allows the resistance of a constriction to be
precisely controlled, the mobility of gold at room temperature is
high enough to break the wire completely, resulting in two separate
electrodes, even without applying a bias. Figure \ref{graph:fig1}(b)-(d)
shows the conductance versus time of three wires at room temperature
when initially narrowed by active breaking to $4\,$k$\Omega$, $2\,$k$\Omega$
and $900\,\Omega$. Here, we plot the numerical $\frac{dI}{dV}$ around
zero bias obtained by sweeping the applied voltage from $-100\,$mV
to $100\,$mV while measuring the current through the constriction,
at a rate of $1\,$sweep/s. The sample in Figure \ref{graph:fig1}(e)
was narrowed first to $700\,\Omega$ and then measured with a lock-in
amplifier using an oscillation of $1\,$mV$_{RMS}$ around zero bias.
To emphasize that self-breaking occurs even in the absence of an applied
electric field, we have carried out several hundreds of experiments
in which no bias is applied after narrowing, and have observed self-breaking
in all cases. 

In each sample, the resulting reduction in conductance is not continuous
but evolves in discrete steps. Configurations may persist for long
times before changing, and the time to reach less than one conductance
quantum may vary between tens of minutes, to tens of hours. The rate
of conductance drop until $2\,$G$_{0}$ is $\sim2000\,$s/G$_{0}$
for (b) and (c), $8700\,$s/G$_{0}$ for (d), and $1000\,$s/G$_{0}$
for (e). We note that, as the junction approaches a conductance of
$\sim1\,$G$_{0}$, the junctions are typically only stable for a
few seconds. This behavior corresponds to the removal of the final
atom between the two electrodes until charge can only be transferred
by tunneling processes.

\begin{figure}
\begin{centering}\includegraphics[clip,scale=0.6]{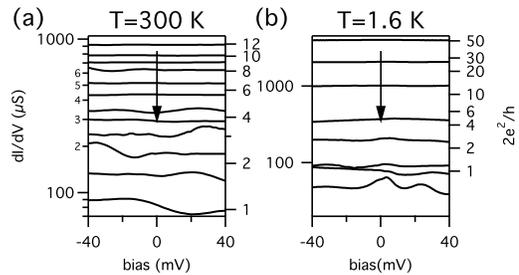}\par\end{centering}

\caption{Differential conductance curves for constrictions of varying conductance
as a function of bias voltage, at room temperature and low temperature.
The room temperature data is recorded during the breaking process.
For the low-temperature data each curve represents a different configuration
obtained with active breaking. (Arrows indicate the sequence in which
the data was collected).}

\label{graph:fig2}
\end{figure}

Despite the unstable nature of the few-atom constrictions at room
temperature, fast acquisition allows $I-V$s to be recorded as the
system evolves. Figure \ref{graph:fig2}(a) shows a selection of the
$dI/dV$ vs $V$ curves collected during self-breaking of the sample
presented in Figure \ref{graph:fig1}(c) at $300\,$K, computed from
the $I-V$ curves as previously described. As we have demonstrated
above, wires narrowed at room temperature are unstable, further characterization
may be performed at low temperature. In contrast to room temperature,
gold wires narrowed by active breaking at $1.6\,$K are stable, and
their $I-V$ characteristics do not vary in time. In this conductance
range, constrictions of a specific target resistance can be created
by actively-controlled electromigration to within $5$\%, even in
the conductance range near $1\,$G$_{0}$. The low-temperature data
in Figure \ref{graph:fig2}(b) was collected with a lock-in amplifier
in the order indicated by the arrow. We see that, in both the unstable
room temperature constrictions, and in the stable low temperature
constrictions, the $\frac{dI}{dV}-V$ curves are flat for $\frac{dI}{dV}\gtrsim5\,$G$_{0}$,
and become highly non-linear near $1\,$G$_{0}$. In addition, measurements
of the samples narrowed at low temperature to $\frac{dI}{dV}\sim1\,$G$_{0}$
at $1.6\,$K reveal negligible magnetic field dependence up to $10\,$T.
For electron interference to be the origin of the observed oscillations
loops of $20\,$nm or less are implied \cite{interference}.

Finally, we present statistics on samples broken at room temperature
with single organic molecules, either by breaking the wires with a
self-assembled monolayer, or by breaking directly in a solution of
the molecules of interest, collected over approximately a year \cite{molecule_list}.
After breaking, samples were cooled to $1.6\,$K and the current was
measured as a function of gate and bias for each device. Of $162$
samples broken by narrowing the constriction to a few atoms wide and
then allowed to self-break with no bias applied, $24$ showed gate
dependence, and none showed indications of transport through gold
clusters (see below). In contrast, of $171$ samples broken by active
breaking into the tunneling regime (applying above-threshold biases
until $>100\,$k$\Omega$), $38$ show gate dependence, of which $6$
samples show indications of transport through gold clusters. 

In discriminating between transport through gold clusters and transport
through single molecules we identify two parameters of a double-tunneling
system with a single charging island: the electron addition energy
and the coupling to the gate electrode \cite{likharev}. We attribute
charging energies less than $100\,$meV in combination with a gate
coupling greater than $0.2$ to gold grains. We reach these parameters
as follows: by identifying single-molecule samples through their vibrational
spectra \cite{osorio} it has been established that, for the molecules
studied here, electron addition energies typically lie in a range
greater than $100\,$meV, comparable with electron addition energies
measured using other techniques \cite{kubatkin}. In addition, transport
that is due to single molecules has consistently shown a low gate
coupling, not larger than $0.15$. Gold clusters, on the the other
hand, would be in direct contact with the gate dielectric, and so
would be expected to have a higher gate coupling, which we typically
find to be $0.25$ \cite{zant_bolotin}.

We have also observed self-breaking in narrowed gold wires on silicon
oxide substrates, instead of aluminium gates, revealing comparable
results to those presented here. In addition, experiments with platinum
wires have shown no self-breaking effect, indicating that the effect
is connected with some intrinsic property of the wire material. The
self-breaking of narrowed wires can be compared with diffusion of
gold, as observed in scanning tunneling microscope (STM) experiments
\cite{roberts}. Here the surface diffusion velocities were measured
in the range from $0.03$ to $0.2\,$\AA/s. Recalculating this as
the time required to jump the nearest neighbour distance $2.88\,$\AA,
surface mobility implies that it takes $10-100\,$s for gold atoms
to move in and out of a contact, contributing to or reducing the conductance
by G$_{0}$. This is within an order of magnitude faster than the
observed rate shown in Figure \ref{graph:fig1} for conductances $>2\,$G$_{0}$,
and an order of magnitude slower than for junctions with conductance
$<2\,$G$_{0}$. Given these variations, there is rough agreement
between this crude calculation and the measurements which seems to
suggest that a diffusion mechanism drives the self-breaking process.

\begin{figure}
\begin{centering}\includegraphics[clip,scale=0.6]{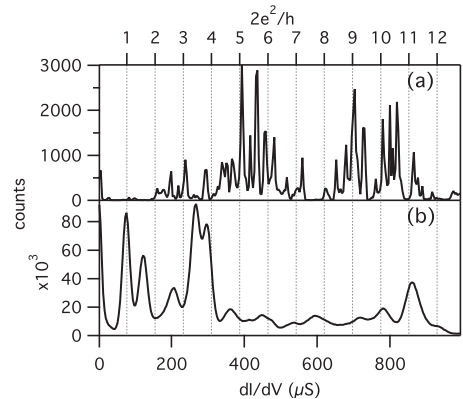}\par\end{centering}

\caption{(a) Conductance histogram constructed from data presented in Figure
\ref{graph:fig1}; (b) conductance histogram constructed from 8 samples
measured with a sampling rate of $100\,\mu$s, broken under a applied
potential of $225\,$mV; each graph uses $20\:$bins per G$_{0}$.}

\label{graph:fig3}
\end{figure}

The traces in Figure \ref{graph:fig1}(b) and (c) seem to imply a
preference for conductances at integer multiples of G$_{0}$, consistent
with previous observations \cite{strachen_houck}. Figure \ref{graph:fig3}(a)
shows a histogram in which the occurance of conductances is plotted
in a histogram, using the data presented in Figures \ref{graph:fig1}(b)-(d).
Figure \ref{graph:fig3}(b) shows a similar histogram of conductance
values built from $8$ samples with a driving field close to the electromigration
threshold, using a much higher sampling rate. It is clear that, while
some of the conductance peaks fall on integer multiples of G$_{0}$,
many others do not, and it is therefore difficult to conclude that
preferred conductances exist. For comparison, STM \cite{brandbyge}
and MCBJ experiments show that at least $\sim230$ traces have to
be considered, implying that the statistics in our experiments may
be too low. In addition, in such experiments only the first three
conductance peaks are visible, for which we obtain fewer statistics.

The cause of the non-linear $\frac{dI}{dV}$ vs $V$ curves shown
in Figure \ref{graph:fig2} is unclear. An explanation for the effect
may come from electron interference in ballistic atomic-scale junctions
\cite{ludoph_untiedt}, which may be of the same order of magnitude
as the measurements presented here. Further measurements, at room
and low temperatures, have not revealed a clear suppression of the
non-linearities near integer multiples of G$_{0}$, which would strongly
corroborate the electron interference theory. An alternative cause
could stem from contamination in the junction. STM experiments establish
that contamination by air (notably water vapour) both modifies the
surface of gold samples \cite{roberts} and changes the second derivative
G$''$ of one and two atom junctions \cite{hansen}. We believe that
this is unlikely to explain the variation in G$''$ observed in our
junctions, which is $\sim1000$ times higher than reported values
(reference 17).

Financial support was obtained from the Dutch organization for Fundamental
Research on Matter (FOM), and the `Nederlandse Organisatie voor Wetenschappelijk
Onderzoek' (NWO). We thank M. Poot for the data on silicon oxide substrates.
K.O'N. was supported by the Marie Curie Fellowship organization. The
authors are grateful to A. Ito, T. Bjørnholm and M. Ruben for providing
the molecules.

\newpage{}
\end{document}